\documentclass{nature_kcp}
\usepackage{graphicx}

\newcommand{\ltappeq}{\raisebox{-0.6ex}{$\,\stackrel{\raisebox{-.2ex}{$\textstyle <$}}{\sim}\,$}}
\newcommand{\gtappeq}{\raisebox{-0.6ex}{$\,\stackrel{\raisebox{-.2ex}{$\textstyle >$}}{\sim}\,$}}

\def\spose#1{\hbox to 0pt{#1\hss}}
\def\lta{\mathrel{\spose{\lower 3pt\hbox{$\mathchar"218$}} \raise 2.0pt\hbox{$\mathchar"13C$}}}
\def\gta{\mathrel{\spose{\lower 3pt\hbox{$\mathchar"218$}} \raise 2.0pt\hbox{$\mathchar"13E$}}}
\newcommand{\apss}{\textit{Astrophys. Space Sci.}}

\newcommand{\apj}{\textit{Astrophys. J.}}
\newcommand{\aj}{\textit{Astronom. J.}}
\newcommand{\apjl}{\textit{Astrophys. J. Lett.}}

\newcommand{\mnras}{\textit{Mon. Not. R. Astron. Soc.}}

\newcommand{\aap}{\textit{Astron. \& Astronphys.}}
\newcommand{\aaps}{\textit{Astron. \& Astronphys. Supp.}}
\newcommand{\nat}{\textit{Nature}}

\title{\Large \bf Two Populations of X-ray Pulsars Produced by Two Types of Supernovae}
 
\author{Christian Knigge$^{1}$, Malcolm J. Coe$^1$ \& Philipp Podsiadlowski$^2$}

\begin{document}

\topmargin=0.7in

\spacing{1.0}
\maketitle

\begin{affiliations}
 \item University of Southampton, School of Physics and Astronomy,
   Southampton SO17 1BJ, UK 
 \item University of Oxford, Department of Physics, Oxford OX1 3RH, UK
 \end{affiliations}

\begin{abstract}
Two types of supernova are thought to produce the overwhelming majority of
neutron stars in the Universe\cite{2003ApJ...591..288H}.
The first type, iron-core collapse
supernovae, occurs when a high-mass star develops a degenerate iron
core that exceeds the Chandrasekhar limit\cite{2005NatPh...1..147W}. The second type, 
electron-capture supernovae, is associated with the collapse of a
lower-mass oxygen-neon-magnesium core as it loses 
pressure support owing to the sudden capture of electrons by neon
and/or magnesium
nuclei\cite{1984ApJ...277..791N,1987ApJ...322..206N}. It has hitherto
been impossible to 
identify the two distinct families of neutron stars produced in these
formation channels. Here we report that a large,
well-known class of neutron-star-hosting X-ray 
pulsars is actually composed of two distinct sub-populations with different
characteristic spin periods, orbital periods and orbital
eccentricities. This class, the Be/X-ray binaries, contains
neutron stars that accrete material from a more massive companion
star\cite{2011Ap&SS.332....1R}. 
The two sub-populations are most probably associated with the two
distinct types of neutron-star-forming supernovae, with electron-capture
supernovae preferentially producing system with short spin period,
short orbital periods and low eccentricity. Intriguingly, the split
between the two sub-populations is clearest in the distribution of the
logarithm of spin period, a result that had not been predicted and
which still remains to be explained.

\end{abstract}

Be/X-ray binaries (BeXs) are strong X-ray sources because their
neutron stars accrete material at a relatively high rate. Their
mass-losing Be-type companions are fast-rotating $8 M_{\odot} - 18  
M_{\odot}$ main-sequence stars that are surrounded by 
circumstellar ``decretion disks''. These disks are fueled by the
injection of mass and angular momentum at the stellar
surface\cite{1991MNRAS.250..432L}. Neutron star spin periods in BeXs 
are typically 1 - 1,000~s, and BeX orbits are usually
elliptical, with orbital periods ranging from about 10 - 1,000~d. Most
of the accretion takes place during periastron passages,
when the neutron star passes close to, or even through, the Be-star
decretion disk. 

BeXs are exceptionally abundant in the Small Magellanic Cloud (SMC),
where a burst of star formation $\simeq$60 Myr
ago\cite{2004AJ....127.1531H} seems to have produced a large
population of these systems\cite{2004A&A...414..667H,2005MNRAS.356..502C}. 
In fact, the SMC contains a comparable number of BeXs to the Milky Way (MW), 
even though the mass ratio  of the two galaxies is about 
1:100. By contrast,
the number of BeXs in the Large Magellanic Cloud
(LMC) is broadly in line with its stellar mass content when compared
with the Milky Way.

In the context of studying neutron star formation channels, it is useful to
focus on well-defined, simple and ``clean'' populations of neutron-star-hosting
systems (that is, systems in which the orbital parameters have not yet
evolved since the supernova) that nevertheless span a wide range of
properties. BeXs can 
provide this. This is not only because the neutron stars in BeXs all 
have the same type of companion, but also because the accretion
process itself seeems to be universal, with the neutron star spin in or near an
equilibrium state in which the magnetospheric radius of the neutron star equals
the Keplerian co-rotation
radius\cite{1989A&A...223..196W,1996A&A...314L..13L,2003IAUS..214..215L}. 
This conclusion is suggested empirically by the location of 
BeXs in the $\log(P_{\rm orb})$-$\log(P_{\rm spin})$ plane (the Corbet
diagram\cite{1984A&A...141...91C};$P_{\rm orb}$ and $P_{\rm spin}$ are
the orbital and period period, respectively), where they tend to lie
along a line with slope $\alpha \simeq 2$ (Fig.~1). 

The correlation in Fig.~1 between $P_{\rm spin}$ and $P_{\rm orb}$
among BeXs is highly significant, but the data have large scatter. 
Despite this scatter, however, the 1-D projections of the data 
onto the $\log{P_{\rm orb}}$ and $\log{P_{\rm spin}}$ axes both suggest that
the BeX population might be bimodal. More specifically, the two
sub-populations suggested by the data in Fig.~1 have characteristic
periods of $P_{\rm orb} \simeq 40$~d and $P_{\rm spin} \simeq 10$~s (short-period
mode), and $P_{\rm orb} \simeq 100$~d and $P_{\rm spin} \simeq 200$~s (long-period
mode). The bimodality of the BeX population seems to be more
prominent in $\log{P_{\rm spin}}$ than in $\log{P_{\rm orb}}$. This is
helpful, because there are additional BeXs, not shown in Fig.~1,
for which $P_{\rm spin}$ is known, but $P_{\rm orb}$ is not. For the 
purpose of analysing the $P_{\rm spin}$ data on their own, we can
therefore add these systems to the list of confirmed and probable BeXs.

Such an analysis is shown in Fig.~2. It confirms that the
$\log{P_{\rm spin}}$ distribution of BeXs contains two distinct
sub-populations with characteristic spin periods of 
$P_{\rm spin} \simeq 10$~s and $P_{\rm spin} \simeq 200$~s and 
similar dispersions of about $0.4$ dex. The short-$P_{\rm spin}$ and  long-$P_{\rm
spin}$ sub-populations contribute about 35 and 65\% to the total
number, respectively. The split into these two
sub-populations is highly statistically significant for the full
sample, and remains significant even if the data set is divided by 
host galaxy. The split also remains significant if we consider only
spectroscopically confirmed BeXs. Finally, the evidence for two
sub-populations even remains significant if we use non-parametric
statistical tests (which are less powerful, but more robust than the
KMM algorithm; see the {\em Supplementary Information} for details).

As shown explicitly in Fig~2 (bottom panel), the
double-Gaussian decomposition of the independent SMC and MW+LMC
samples are consistent with each other. This makes it highly unlikely that
selection effects are responsible for the observed bimodality.
In any case, it is hard to conceive of a selection bias that
would select specifically against BeXs with intermediate
$P_{\rm spin}$ and/or $P_{\rm orb}$. We therefore believe that the two modes
of the $\log(P_{\rm spin})$ distribution correspond to physically
distinct BeX sub-populations.

In principle, there are at least three ways to account for the
existence of these sub-populations. First, they could correspond to
two distinct neutron star spin equilibria that are accessible at all orbital
periods. However, even though the bimodality is stronger in
$P_{\rm spin}$ than in $P_{\rm orb}$, the existence of the $P_{\rm spin} - P_{\rm orb}$
correlation effectively rules out this possibility. 
Second, $P_{\rm orb}$ might be time-dependent for BeXs, with the two
sub-populations representing two distinct long-lived evolutionary
stages. However, the time scale for stellar-wind-driven changes in $P_{\rm orb}$ 
in the BeX phase, $\tau_{\rm P_{orb}} \sim 100~- 1000~{\rm
 Myr}$ (refs 14, 15)\nocite{1988A&AS...72..259D,1991MNRAS.253....9T}, is
substantially longer than the maximum duration of this phase,
$\tau_{\rm BeX,max} \sim 20~{\rm Myr}$, the lifetime of an $8
M_{\odot}$ star. Thus $P_{\rm orb}$ evolution also cannot account for
the two observed sub-populations.

The third explanation that we consider is that the two sub-populations
represent two distinct BeX formation channels. The most obvious
possibility, with the farthest reaching implications, is that
the two channels are associated with the two distinct types of supernova
event noted above. More specifically, an iron-core-collapse supernova marks the end point of
the evolution of any sufficiently massive star, whereas an electron-capture supernova can
occur only under highly restrictive conditions. In 
particular, an electron-capture supernova requires that the core reaches the critical
density for electron-capture to occur, $4.5\times 10^9 {\rm
g~cm^{3}}$ (ref. 16)\nocite{2005MNRAS.361.1243P}. These conditions
might be met in the late evolution of intermediate-mass
stars\cite{1984ApJ...277..791N,1987ApJ...322..206N} (those with
initial masses satisfying $8 M_{\odot} \ltappeq M_{\rm init}
\ltappeq 10 M_{\odot}$), although the relevant mass range is uncertain and may be 
quite small\cite{2004ApJ...612.1044P,2008ApJ...675..614P}. However, it is much
easier to meet the conditions for electron-capture supernovae naturally in binary
systems\cite{2004ApJ...612.1044P}.

The outcome of electron-capture supernovae differs from that of iron-core-collapse supernovae in two fundamental 
ways. First, electron-capture supernovae should produce somewhat less massive neutron stars
($\ltappeq 1.3~M_{\odot}$) than iron-core-collapse supernova
($1.4~M_{\odot}$) (ref. 3)\nocite{1984ApJ...277..791N}. 
Second, electron-capture supernovae are expected to impart much 
smaller kicks to the neutron stars they produce (average kick velocity
of $\ltappeq 50~{\rm km s^{-1}}$) than are iron-core-collapse
supernovae ($\gtappeq 200~{\rm km s^{-1}}$)(ref. 17)\nocite{2004ApJ...612.1044P}. 
In binary systems, where kicks induce 
orbital eccentricity, these differences could naturally give rise to
two distinct sub-populations. The more conventional iron-core-collapse channel would
produce high-eccentricity binaries containing high-mass neutron stars,
and the electron-capture channel would produce low-eccentricity binaries containing low-mass
neutron stars\cite{2004ESASP.552..185V,2004ApJ...612.1044P,2005MNRAS.361.1243P,2010ApJ... 719..722S}.

If this is the correct explanation for the two BeX populations we have
discovered, they should differ not only in $P_{\rm spin}$ and $P_{\rm orb}$,
but also in their characteristic orbital eccentricities, $e$. These
have so far been measured for only about 20 BeXs. Figure~3 shows the
distribution of these systems in the $\log{P_{\rm spin}}-e$ plane. Even
though there are only 8 such BeXs with $P_{\rm spin} \gtappeq 40$~s,
it seems that 
long-$P_{\rm spin}$ systems have preferentially higher eccentricities than
short-$P_{\rm spin}$ systems. The figure also shows the 
cumulative eccentricity distributions of the slow ($P_{\rm spin} 
< 40$~s) and fast ($P_{\rm spin} > 40$~s) BeX pulsar
sub-populations. A Kolmogorov-Smirnov test shows that, 
despite the small number of
systems for which eccentricity has been measured, the maximum difference between
these distributions is marginally significant {\em Supplementary Information}). 

The two BeX populations that we have discovered are more clearly separated in
$P_{\rm spin}$ than in $P_{\rm orb}$. Given that $P_{\rm orb}$ does not evolve
significantly within the BeX phase, whereas $P_{\rm spin}$ does, $P_{\rm orb}$ might
be expected $P_{\rm orb}$ to be the more faithful tracer of the 
formation channels. However, the $P_{\rm orb}$ distribution after the
supernova must depend strongly on the $P_{\rm orb}$ distribution
before the supernova,  and both low- and high-velocity kicks can in
principle produce a wide range of post-supernova orbital
periods\cite{2004ApJ...612.1044P}. This may explain why the bimodality
is only marginally observed in the $P_{\rm orb}$ distribution. By
contrast, the equilibrium spin period is expected to depend on 
several system parameters other than
$P_{porb}$ (refs 10-12)\nocite{1989A&A...223..196W,1996A&A...314L..13L,2003IAUS..214..215L}. If
any of these parameters differ systematically between BeXs produced by
the iron-core-collapse and electron-capture channels, $P_{\rm spin}$
may be a more reliable indicator of formation channel than
$P_{\rm orb}$.

Our results suggest numerous avenues for further research. First, 
it is important to expand the data base of BeXs with
reliable measurements of $P_{\rm spin}$, $P_{\rm orb}$ and $e$, 
to confirm and further quantify our findings. Second, if short-$P_{\rm spin}$
BeXs are formed by low-kick-velocity electron-capture supernovae, they should have
systematically smaller space velocities than long-$P_{\rm spin}$ BeXs. This
prediction might be testable\cite{2005MNRAS.358.1379C,2010ApJ...716L.140A}. 
Third, short-$P_{\rm spin}$ systems 
should also have systematically lower neutron star masses than 
long-$P_{\rm spin}$ systems
\cite{2004ESASP.552..185V,2004ApJ...612.1044P,2005MNRAS.361.1243P,2010ApJ...719..722S}. This
prediction might also be testable\cite{2010MNRAS.401..252C}. 
Fourth, although our discovery of two populations of Be/X-ray
pulsars is robust, our suggested explanation for their origin is
clearly speculative and demands a fuller investigation. Intriguingly,
recent binary population synthesis calculations have shown that the 
electron-capture supernova channel may be very efficient at forming
BeXs\cite{2009ApJ...699.1573L}. However, whereas the same population synthesis
models also suggest that the electron-capture channel accounts for the
overabundance of BeXs in the SMC\cite{2009ApJ...699.1573L}, the
two BeX populations we have discovered appear to have similar
relative abundances in the SMC and the Milky Way.

\clearpage

\bibliographystyle{naturemag}

\begin{addendum}
 \item[Acknowledgements] Research support for this project was
   provided by the UK's Science and  
   Technology Facilities Council. The authors would like to thank Tom
   Maccarone and Tim Linden for useful 
   discussions and Leslie Sage for prompting us to take a quantitative
   look at the eccentricity data. 

\item[Author Contributions] C.K. carried out the statistical analysis
  for this project and wrote most of the text. M.J.C. compiled the
  high-mass X-ray binary data set that forms the basis for our
  analysis and collaborated with C.K. on all aspects of the project
  from its inception. P.P. was instrumental to the theoretical
  interpretation of the results and also contributed to the final text. 
  All authors discussed the results and their presentation.

\item[Author Information] Reprints and permissions information is
   available at \\ npg.nature.com/reprintsandpermissions. The authors
   declare that they have no competing financial interests. Correspondence
   and requests for materials should be addressed to C. K. ~(email:
   C.Knigge@soton.ac.uk).
\end{addendum}

\clearpage
\thispagestyle{empty}
\begin{figure}
\begin{center}
\includegraphics[scale=0.58, angle=0]{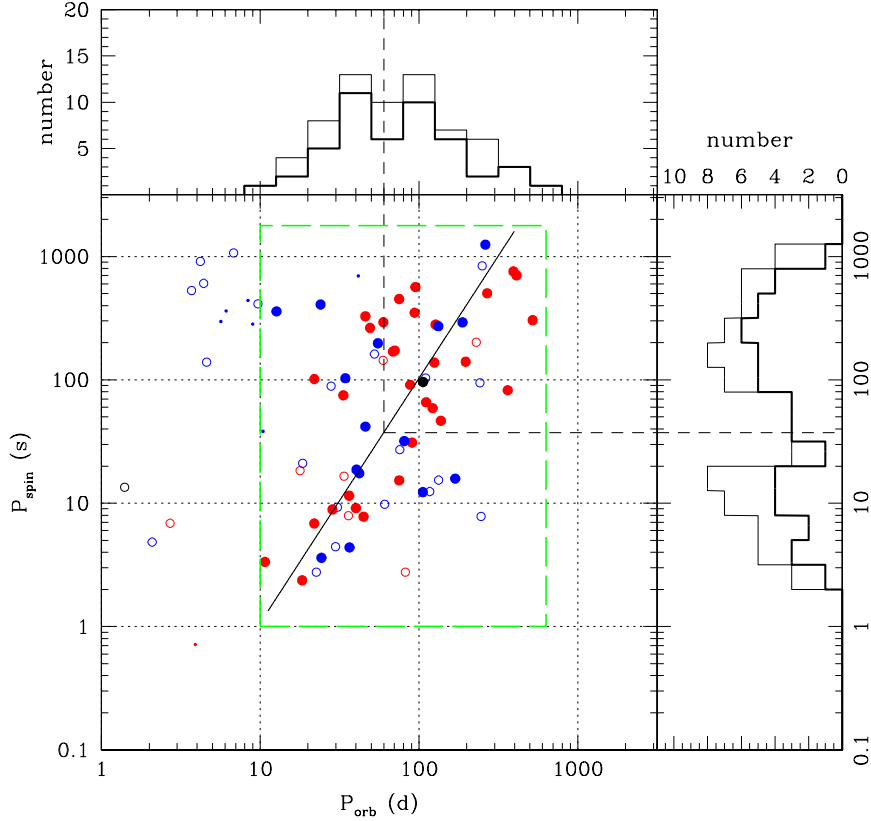}
\end{center}
\caption[]{\small {\bf The Corbet diagram for high-mass X-ray binaries.} The 
central panel shows $\log{P_{\rm orb}}$ versus $\log{P_{\rm spin}}$ for
neutron-star-hosting high-mass X-ray binaries. Filled circles correspond to spectroscopically
confirmed BeXs, small dots to confirmed non-BeX systems and open circles
to candidate BeXs. There are additional confirmed and candidate BeXs for
which only $P_{\rm orb}$ or $P_{\rm spin}$ is known, but these are not
shown. The dashed green lines mark a selection box that conservatively
includes all confirmed BeXs for which $P_{\rm orb}$ and
$P_{\rm spin}$ have been measured. Candidate systems outside this box
are excluded from our sample of probable BeXs. The spin 
and orbital periods of confirmed and probable BeX systems are
correlated. The Spearman-rank correlation coefficient is $\rho = 0.49$ ($p =
3\times10^{-5}$; $N = 66$) 
for the full sample and $\rho = 0.49$ ($p = 4 \times 10^{-4}$; $N = 47$) 
for the confirmed systems (see {\em Supplementary Information} for
a definition of $p$-values). The scatter around the 
correlation is $\sigma_{\rm \log{P_{\rm spin}}} = 0.7$~dex 
relative to the best-fitting line with slope $\alpha
= 2$ (solid black line). Different colours indicate different host
galaxies: blue, Milky Way; red, MC; black, LMC. The histograms shown in the
top and right-hand panels show the number of BeXs with spin and
orbital periods in the respective ranges covered by the selection
box. In each of these panels, the thick line corresponds to
confirmed BeXs only, and the thin line corresponds 
to confirmed and probable BeXs. The vertical dashed line is drawn at
$P_{\rm orb} = 60$~d, the location of the apparent dip in the
$\log{P_{\rm orb}}$ distribution. This value of $P_{\rm orb}$ corresponds to $P_{\rm spin}
\simeq 40$~s (horizontal dashed line), which marks a more pronounced 
dip in the $\log{P_{\rm spin}}$ distribution.}
\label{fig1}
\end{figure}

\clearpage
\spacing{1}
\thispagestyle{empty}
\begin{figure}
\begin{center}
\includegraphics[scale=0.59, angle=0]{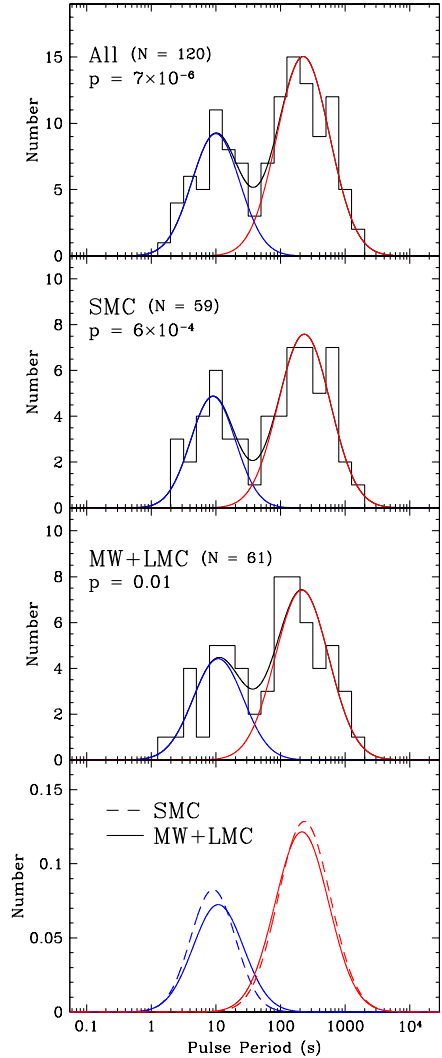}
\end{center}
\caption[]{\small {\bf The $\log{P_{\rm spin}}$ distribution of confirmed and probable
BeXs.} {\bf Top panel:} Distribution for all systems. {\bf Middle panels:} Distribution broken down by
host galaxy: SMC (second from top); Milky Way (MW)+LMC (third from top). All of 
these distributions are bimodal, and the double-Gaussian decomposition 
suggested by the KMM algorithm\cite{1994AJ....108.2348A} is shown
in each panel. The number of systems contributing to each
observed distribution and the associated $p$-value provided by the
algorithm are also shown. Applying the KMM test to the subset of spectroscopically
confirmed systems (not shown) gives $p = 8\times10^{-3}$ (N = 64).
{\bf Bottom panel:} Direct comparison of the
decompositions for the independent SMC and Milky Way+LMC
populations, showing them to be mutually consistent. Additional
details regarding the statistical evidence for the existence of 
distinct sub-populations in the $P_{\rm spin}$ data are given in {\em
Supplementary Information}.}
\label{fig2}
\end{figure}

\clearpage
\spacing{1}
\thispagestyle{empty}
\begin{figure}
\begin{center}
\includegraphics[scale=0.59,angle=270]{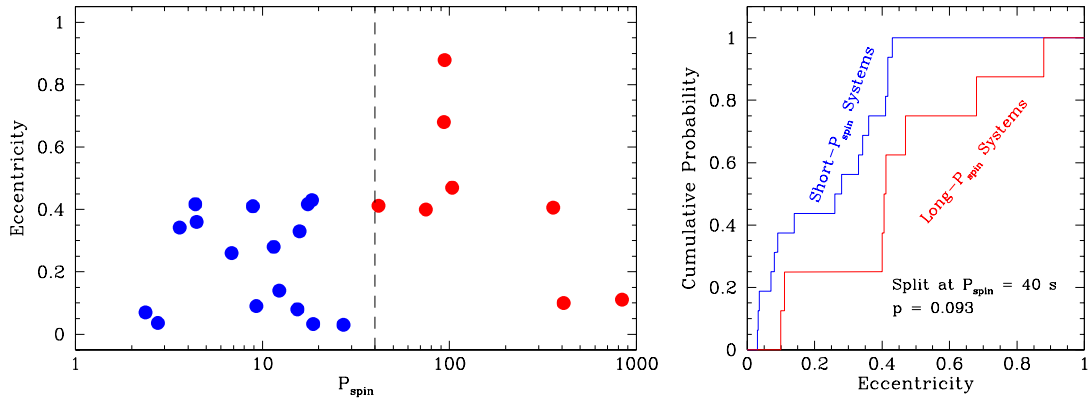}
\end{center}
\caption[]{\small{\bf The dependence of eccentricity on $P_{\rm spin}$ among
BeXs.} {\bf Left panel:} $P_{\rm spin}$ versus eccentricity for all confirmed
and probable BeXs with measured spin periods and eccentricities. The
vertical dashed line 
marks the approximate division between the short-$P_{\rm spin}$ and
long-$P_{\rm spin}$ sub-populations (Figs~1 and 2). {\bf Right
panel:} Cumulative eccentricity distributions of these two
sub-populations. A Kolmogorov-Smirnov test provides marginal evidence
for a difference between these distributions ($p = 0.093$), 
with the short-$P_{\rm spin}$ population being characterized by
lower eccentricities (see {\em Supplementary Information} for
additional discussion).}
\label{fig3}
\end{figure}

\clearpage

\setcounter{figure}{0}

\begin{flushleft}
\title{\bf \Large Supplementary Information}
\end{flushleft} 

\topmargin=0.7in

\spacing{1.0}
\maketitle



\section{Introduction} 

In the main paper, we present evidence for the existence of two
distinct sub-populations of Be/X-ray binaries (BeXs), which are most
clearly separated in the $\log{P_{\rm spin}}$ distribution. We argue that
these two populations are most likely associated with different types
of neutron star formation channels, with electron-capture
supernovae primarily producing short-$P_{\rm spin}$ systems, and
iron-core collapse supernovae primarily producing
long-$P_{\rm spin}$ systems. This particular assignment is largely based
on the eccentricity distribution of BeXs, which suggests that
short-$P_{\rm spin}$ (long-$P_{\rm spin}$) systems are preferentially
associated with low-$e$ (high-$e$) binary orbits. 

Below, we provide supplementary information for several
aspects of our analysis. First, we add some historical
context to our discovery of two BeX sub-populations with
distinct characteristic spin periods. Second, we provide 
additional details regarding the significance with which these
sub-populations are detected in our data. Third, we
discuss some subtleties regarding the interpretation of the BeX 
eccentricity distribution that help to properly assess the strength
of the evidence for a connection between $e$ and $P_{\rm spin}$. We also
show explicitly that there is no evidence for a connection between $e$
and $P_{\rm orb}$.

\section{Historical Context}

The possible existence of a gap in the distribution between 
$P_{\rm spin} \simeq 10$~s and $P_{\rm spin} \simeq 
100$~s was actually first suggested 30 years ago, although only $10 - 20$  
systems were known at the
time\cite{1977Natur.266..123R,1982Ap&SS..81..379B}. This suggestion 
does not seem to have been followed-up, however, perhaps because some
later discoveries fell into this gap. In any case, to the best of
our knowledge, the evidence for multiple sub-populations in the
spin-period distribution has not been seriously reconsidered until
now.

\section{The Statistical Evidence for Two Distinct Populations of BeXs}

We show in the main text that the evidence for two sub-populations
contributing to the $\log{P_{\rm spin}}$ distribution of BeXs is highly
statistically significant according to the standard KMM
test$^{25}$.
The test statistic used by KMM is the
likelihood ratio between the best single-Gaussian representation and
the best double-Gaussian representation of the data. For the latter,
we always find that the variances of the two Gaussians are consistent
with each other. In our application of KMM, we thus adopt equal
variances in the double Gaussian representation and then calculate the
$p$-values robustly from bootstrap simulations\cite{2010ApJ...718.1266M}. 
Here and throughout this paper, the quoted $p$-values have the usual
statistical meaning, i.e. they 
represent the probability of obtaining a test statistic at least as
extreme as the observed one, under the assumption that the null
hypothesis is correct. In the case of KMM, the null hypothesis is that
the data are drawn from a single Gaussian.

Figure~2 in the main text shows that the best double-Gaussian
representation actually provides quite a good match to the
$\log{P_{\rm spin}}$ data. It is nevertheless interesting to ask if we can
find evidence for multiple sub-populations in our data set even
without the assumption of normality. This can be checked by
non-parametric methods, such as the ``bandwidth test''\cite{silverman}.
The standard bandwidth test is a non-parametric test for
multi-modality based on bootstrapping from  
a kernel density estimate of the underlying probability
distribution function (PDF). The statistic used in the 
test is the smallest kernel width that results in a unimodal
estimate of the PDF. The standard bandwidth test is a strict
test of modality, in the sense that the null hypothesis is only that the
underlying PDF is unimodal, with no additional assumptions concerning
its shape. However, a unimodal distribution with an obvious
``shoulder'', for example, may still be considered to provide 
strong evidence for
multiple sub-populations. We can therefore also define an
alternative test, in which the statistic used is the smallest kernel
width that results in a kernel density estimate of the PDF that has only two
inflection points\cite{convex}. Being entirely non-parametric, 
both of these tests are extremely robust. However, they
are also much less powerful than KMM, mainly in the sense that
their null hypotheses are so general that they are hard to rule
out. The resulting $p$-values, which can be calibrated via Monte-Carlo
simulations\cite{hall}, are thus always larger than those provided by
KMM.

We have applied both versions of the bandwidth test to the $P_{\rm spin}$
data for confirmed and probable BeXs. For the standard (modality) 
bandwidth test, we first apply a skew-minimising Box-Cox transformation to
the $\log{P_{\rm spin}}$ data, because modes associated with distinct
sub-populations can be either emphasised or suppressed by certain 
transformations\cite{2008MNRAS.386.1426K}. A transformation to
symmetry is expected to make any inherent bimodality (strictly
speaking: ``bimodalizability'') easier to detect\cite{wys}. 
For our BeX data set, the classic bandwidth test
then rejects the null hypothesis of unimodality with $p =
9 \times 10^{-3}$ 
(application of this test directly to the $\log{P_{\rm spin}}$ data
would give $p = 0.04$). Applying the alternative (inflection point)
bandwidth test directly to the $\log{P_{\rm spin}}$ data  
rejects the null hypothesis that there are only two inflection points
in the underlying distribution with $p = 1 \times 10^{-3}$
(application of this test to the deskewed data would give $p = 5
\times 10^{-3}$). Given the robust non-parametric nature of these
tests, we consider these results to be strong evidence for the
existence of  two distinct BeX sub-populations.

\section{The Eccentricity Distribution of BeXs}

Figure~3 in the main text shows that there is marginally significant
evidence that the eccentricity distribution of long-$P_{\rm spin}$ BeXs is
different from that of short-$P_{\rm spin}$ systems. Some care has to be
taken in interpreting this result. 

First, tidal circularization will always tend to reduce the
eccentricity of close 
binary systems. However, the orbital periods within our BeX sample 
($P_{\rm orb} > 10~{\rm d}$) are too long for this mechanism to be
effective\cite{2003A&A...405..677N}. Second, other things being equal,
a given supernova kick velocity will induce 
larger orbital eccentricities in wide, long-period, binaries, 
since such systems are more weakly bound. Given that
$P_{\rm spin}$ is correlated with $P_{\rm orb}$ in BeXs, this could
produce a $P_{\rm spin} - e$ correlation in the observed sense.

Does this latter bias affect our analysis? Supplementary Figure~1
shows the $P_{\rm orb}$ versus $e$ distribution for our BeX
sample. Superposed on this, we also show the {\em predicted} 
location\cite{1995MNRAS.274..461B}, under
the assumption that they all receive the large kicks typically
associated with conventional iron-core-collapse  
($\left<v_{\rm kick} \right>\simeq 450~{\rm km~s^{-1}}$).
Two key conclusions may be drawn immediately from this plot. First, 
the observed distribution is clearly inconsistent with being
drawn from a single, high-kick population. In fact, the low
eccentricities ($e \ltappeq 0.4$) seen in many BeXs with $P_{\rm orb}
\gtappeq 20~{\rm d}$ cannot be explained by large kicks at
all. A possible connection between such low-eccentricity
HMXBs\cite{2002ApJ...574..364P} and low-kick electron capture
supernovae has indeed been 
suggested before$^{17,19}$.
Second, the $P_{\rm orb}-e$ plane does not separate the two BeX
sub-populations as cleanly as the $P_{\rm spin}-e$ plane. Indeed, if
we split the BeX sample into short-period ($P_{\rm orb} < 60~{\rm d}$) and
long-period ($P_{\rm orb} > 60~{\rm d}$) systems, their cumulative
eccentricity distributions do not differ significantly ($p =
0.35$). Both conclusions indicate that the link between spin period and
eccentricity suggested in the main paper is not simply due to a
$P_{\rm orb}$-dependent bias. 

Finally, there are two obvious
outliers from the trend that long-$P_{\rm spin}$ systems have
preferentially high eccentricities (see Figure~3 in the main
text). These systems are X~Per and 1A~1118-616, which actually have
the longest spin periods ($P_{\rm spin} = 837.7$~s and $407.7$~s,
respectively), but $e \simeq 0.1$. Both of these object are known to be 
highly unusual within the BeX class. For example, X Per
does not show the transient X-ray bursts associated with periastron
passages that are the hallmark of almost all BeXs, while 1A 1118-616
is one of the furthest systems from the standard $P_{\rm orb}-P_{\rm spin}$
correlation (Figure~1 of the main text). Indeed, in both systems, the
dominant accretion mode probably
differs from that which establishes the spin equilibrium in other
BeXs\cite{2001ApJ...546..455D,2011A&A...527A...7S}. This may explain
why $P_{\rm spin}$ is {\em not} a valid tracer of formation channel in
these two objects. Also, since 1A~1118-616 has a surprisingly short orbital
period considering its slow spin ($P_{\rm orb} =
24$~d\cite{2011A&A...527A...7S}), the low eccentricity of 
this system is not actually in terrible conflict with a standard
(high) kick scenario\cite{1995MNRAS.274..461B}. 
If X~Per and 1A~1118-616 are excluded from the
eccentricity distribution of long-$P_{\rm spin}$ systems, the difference 
to the short-$P_{\rm spin}$ systems becomes highly significant ($p =
0.006$). {\em A posteriori} outlier removal is, of course, 
a dangerous statistical practice, so this $p$-value cannot be
taken quite at face value. Nevertheless, it is also interesting to
note that all other high-$P_{spin}$ systems in Figure~3 do, in fact,
lie in the region predicted for systems produced via high-kick supernovae.

\thispagestyle{empty}
\begin{sfigure}
\begin{center}
\includegraphics[scale=0.6,angle=270]{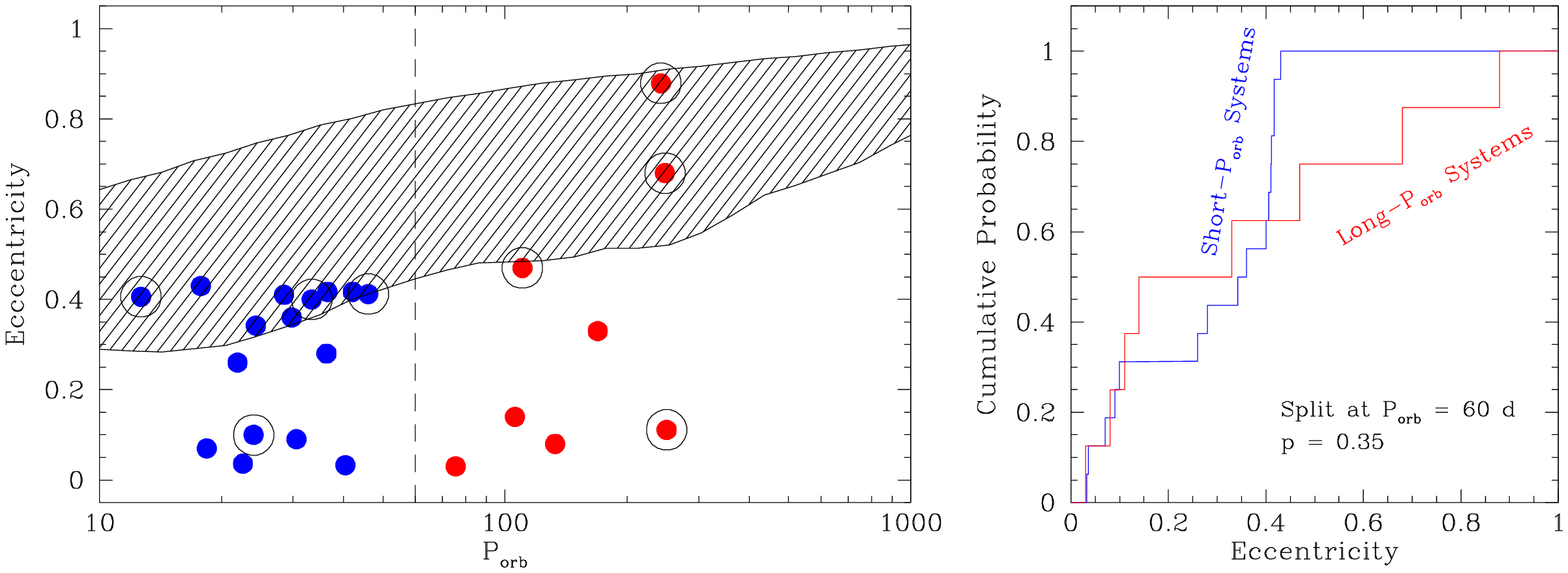}
\end{center}
\caption[]{The dependence of eccentricity on $P_{\rm orb}$ among BeXs. 
The left panel shows $P_{\rm orb}$ versus 
eccentricity for all confirmed and probable 
BeX systems with measured spin and orbital periods and
eccentricities. Long-$P_{spin}$ systems with $P_{spin} > 40$~s are
additionally marked with open circles. The vertical dashed line 
marks the approximate division between the short-$P_{\rm orb}$ and
long-$P_{\rm orb}$ sub-populations (see Figure~1 in the main
text). The shaded area shows the region predicted to contain 60\% of
systems by population synthesis calculations adopting a kick velocity  
distribution appropriate for iron-core-collapse with mean $\left< v_{\rm kick}
\right> \simeq 450~{\rm km~s^{-1}}$\cite{1995MNRAS.274..461B}. The top right
panel shows the cumulative eccentricity distributions of these two
sub-samples. A Kolmogorov-Smirnov test provides no evidence for a 
significant difference between these distributions ($p = 0.35$).} 
\label{supfig1}
\end{sfigure}

\clearpage


\bibliographystyle{naturemag}

\begin{thebibliography}{10}
\expandafter\ifx\csname url\endcsname\relax
  \def\url#1{\texttt{#1}}\fi
\expandafter\ifx\csname urlprefix\endcsname\relax\def\urlprefix{URL }\fi
\providecommand{\bibinfo}[2]{#2}
\providecommand{\eprint}[2][]{\url{#2}}

\bibitem{2003ApJ...591..288H}
\bibinfo{author}{{Heger}, A.}, \bibinfo{author}{{Fryer}, C.~L.},
  \bibinfo{author}{{Woosley}, S.~E.}, \bibinfo{author}{{Langer}, N.} \&
  \bibinfo{author}{{Hartmann}, D.~H.}
\newblock \bibinfo{title}{{How Massive Single Stars End Their Life}}.
\newblock \emph{\bibinfo{journal}{\apj}} \textbf{\bibinfo{volume}{591}},
  \bibinfo{pages}{288--300} (\bibinfo{year}{2003}).

\bibitem{2005NatPh...1..147W}
\bibinfo{author}{{Woosley}, S.} \& \bibinfo{author}{{Janka}, T.}
\newblock \bibinfo{title}{{The physics of core-collapse supernovae}}.
\newblock \emph{\bibinfo{journal}{Nature Physics}}
  \textbf{\bibinfo{volume}{1}}, \bibinfo{pages}{147--154}
  (\bibinfo{year}{2005}).

\bibitem{1984ApJ...277..791N}
\bibinfo{author}{{Nomoto}, K.}
\newblock \bibinfo{title}{{Evolution of 8-10 solar mass stars toward electron
  capture supernovae. I - Formation of electron-degenerate O + NE + MG cores}}.
\newblock \emph{\bibinfo{journal}{\apj}} \textbf{\bibinfo{volume}{277}},
  \bibinfo{pages}{791--805} (\bibinfo{year}{1984}).

\bibitem{1987ApJ...322..206N}
\bibinfo{author}{{Nomoto}, K.}
\newblock \bibinfo{title}{{Evolution of 8-10 solar mass stars toward electron
  capture supernovae. II - Collapse of an O + NE + MG core}}.
\newblock \emph{\bibinfo{journal}{\apj}} \textbf{\bibinfo{volume}{322}},
  \bibinfo{pages}{206--214} (\bibinfo{year}{1987}).

\bibitem{2011Ap&SS.332....1R}
\bibinfo{author}{{Reig}, P.}
\newblock \bibinfo{title}{{Be/X-ray binaries}}.
\newblock \emph{\bibinfo{journal}{\apss}} \textbf{\bibinfo{volume}{332}},
  \bibinfo{pages}{1--29} (\bibinfo{year}{2011}).

\bibitem{1991MNRAS.250..432L}
\bibinfo{author}{{Lee}, U.}, \bibinfo{author}{{Osaki}, Y.} \&
  \bibinfo{author}{{Saio}, H.}
\newblock \bibinfo{title}{{Viscous excretion discs around Be stars}}.
\newblock \emph{\bibinfo{journal}{\mnras}} \textbf{\bibinfo{volume}{250}},
  \bibinfo{pages}{432--437} (\bibinfo{year}{1991}).

\bibitem{2004AJ....127.1531H}
\bibinfo{author}{{Harris}, J.} \& \bibinfo{author}{{Zaritsky}, D.}
\newblock \bibinfo{title}{{The Star Formation History of the Small Magellanic
  Cloud}}.
\newblock \emph{\bibinfo{journal}{\aj}} \textbf{\bibinfo{volume}{127}},
  \bibinfo{pages}{1531--1544} (\bibinfo{year}{2004}).

\bibitem{2004A&A...414..667H}
\bibinfo{author}{{Haberl}, F.} \& \bibinfo{author}{{Pietsch}, W.}
\newblock \bibinfo{title}{{X-ray observations of Be/X-ray binaries in the
  SMC}}.
\newblock \emph{\bibinfo{journal}{\aap}} \textbf{\bibinfo{volume}{414}},
  \bibinfo{pages}{667--676} (\bibinfo{year}{2004}).

\bibitem{2005MNRAS.356..502C}
\bibinfo{author}{{Coe}, M.~J.}, \bibinfo{author}{{Edge}, W.~R.~T.},
  \bibinfo{author}{{Galache}, J.~L.} \& \bibinfo{author}{{McBride}, V.~A.}
\newblock \bibinfo{title}{{Optical properties of Small Magellanic Cloud X-ray
  binaries}}.
\newblock \emph{\bibinfo{journal}{\mnras}} \textbf{\bibinfo{volume}{356}},
  \bibinfo{pages}{502--514} (\bibinfo{year}{2005}).

\bibitem{1989A&A...223..196W}
\bibinfo{author}{{Waters}, L.~B.~F.~M.} \& \bibinfo{author}{{van Kerkwijk},
  M.~H.}
\newblock \bibinfo{title}{{The relation between orbital and spin periods in
  massive X-ray binaries}}.
\newblock \emph{\bibinfo{journal}{\aap}} \textbf{\bibinfo{volume}{223}},
  \bibinfo{pages}{196--206} (\bibinfo{year}{1989}).

\bibitem{1996A&A...314L..13L}
\bibinfo{author}{{Li}, X.} \& \bibinfo{author}{{van den Heuvel}, E.~P.~J.}
\newblock \bibinfo{title}{{On the relation between spin and orbital periods in
  Be/X-ray binaries.}}
\newblock \emph{\bibinfo{journal}{\aap}} \textbf{\bibinfo{volume}{314}},
  \bibinfo{pages}{L13--L16} (\bibinfo{year}{1996}).

\bibitem{2003IAUS..214..215L}
\bibinfo{author}{{Liu}, Q.~Z.}, \bibinfo{author}{{Li}, X.~D.} \&
  \bibinfo{author}{{Wei}, D.~M.}
\newblock \bibinfo{title}{{The Relation between Spin and Orbital Periods in
  HMXBs}}.
\newblock In \bibinfo{editor}{{X.~D.~Li, V.~Trimble, \& Z.~R.~Wang}} (ed.)
  \emph{\bibinfo{booktitle}{High Energy Processes and Phenomena in
  Astrophysics}}, vol. \bibinfo{volume}{214} of \emph{\bibinfo{series}{IAU
  Symposium}}, \bibinfo{pages}{215--217} (\bibinfo{year}{2003}).

\bibitem{1984A&A...141...91C}
\bibinfo{author}{{Corbet}, R.~H.~D.}
\newblock \bibinfo{title}{{Be/neutron star binaries - A relationship between
  orbital period and neutron star spin period}}.
\newblock \emph{\bibinfo{journal}{\aap}} \textbf{\bibinfo{volume}{141}},
  \bibinfo{pages}{91--93} (\bibinfo{year}{1984}).

\bibitem{1988A&AS...72..259D}
\bibinfo{author}{{de Jager}, C.}, \bibinfo{author}{{Nieuwenhuijzen}, H.} \&
  \bibinfo{author}{{van der Hucht}, K.~A.}
\newblock \bibinfo{title}{{Mass loss rates in the Hertzsprung-Russell
  diagram}}.
\newblock \emph{\bibinfo{journal}{\aaps}} \textbf{\bibinfo{volume}{72}},
  \bibinfo{pages}{259--289} (\bibinfo{year}{1988}).

\bibitem{1991MNRAS.253....9T}
\bibinfo{author}{{Tout}, C.~A.} \& \bibinfo{author}{{Hall}, D.~S.}
\newblock \bibinfo{title}{{Wind driven mass transfer in interacting binary
  systems}}.
\newblock \emph{\bibinfo{journal}{\mnras}} \textbf{\bibinfo{volume}{253}},
  \bibinfo{pages}{9--18} (\bibinfo{year}{1991}).

\bibitem{2005MNRAS.361.1243P}
\bibinfo{author}{{Podsiadlowski}, P.} \emph{et~al.}
\newblock \bibinfo{title}{{The double pulsar J0737-3039: testing the neutron
  star equation of state}}.
\newblock \emph{\bibinfo{journal}{\mnras}} \textbf{\bibinfo{volume}{361}},
  \bibinfo{pages}{1243--1249} (\bibinfo{year}{2005}).

\bibitem{2004ApJ...612.1044P}
\bibinfo{author}{{Podsiadlowski}, P.} \emph{et~al.}
\newblock \bibinfo{title}{{The Effects of Binary Evolution on the Dynamics of
  Core Collapse and Neutron Star Kicks}}.
\newblock \emph{\bibinfo{journal}{\apj}} \textbf{\bibinfo{volume}{612}},
  \bibinfo{pages}{1044--1051} (\bibinfo{year}{2004}).

\bibitem{2008ApJ...675..614P}
\bibinfo{author}{{Poelarends}, A.~J.~T.}, \bibinfo{author}{{Herwig}, F.},
  \bibinfo{author}{{Langer}, N.} \& \bibinfo{author}{{Heger}, A.}
\newblock \bibinfo{title}{{The Supernova Channel of Super-AGB Stars}}.
\newblock \emph{\bibinfo{journal}{\apj}} \textbf{\bibinfo{volume}{675}},
  \bibinfo{pages}{614--625} (\bibinfo{year}{2008}).

\bibitem{2004ESASP.552..185V}
\bibinfo{author}{{van den Heuvel}, E.~P.~J.}
\newblock \bibinfo{title}{{X-Ray Binaries and Their Descendants: Binary Radio
  Pulsars; Evidence for Three Classes of Neutron Stars?}}
\newblock In \bibinfo{editor}{{V.~Schoenfelder, G.~Lichti, \& C.~Winkler}}
  (ed.) \emph{\bibinfo{booktitle}{5th INTEGRAL Workshop on the INTEGRAL
  Universe}}, vol. \bibinfo{volume}{552} of \emph{\bibinfo{series}{ESA Special
  Publication}}, \bibinfo{pages}{185--194} (\bibinfo{year}{2004}).

\bibitem{2010ApJ...719..722S}
\bibinfo{author}{{Schwab}, J.}, \bibinfo{author}{{Podsiadlowski}, P.} \&
  \bibinfo{author}{{Rappaport}, S.}
\newblock \bibinfo{title}{{Further Evidence for the Bimodal Distribution of
  Neutron-star Masses}}.
\newblock \emph{\bibinfo{journal}{\apj}} \textbf{\bibinfo{volume}{719}},
  \bibinfo{pages}{722--727} (\bibinfo{year}{2010}).

\bibitem{2005MNRAS.358.1379C}
\bibinfo{author}{{Coe}, M.~J.}
\newblock \bibinfo{title}{{An estimate of the supernova kick velocities for
  high-mass X-ray binaries in the Small Magellanic Cloud}}.
\newblock \emph{\bibinfo{journal}{\mnras}} \textbf{\bibinfo{volume}{358}},
  \bibinfo{pages}{1379--1382} (\bibinfo{year}{2005}).

\bibitem{2010ApJ...716L.140A}
\bibinfo{author}{{Antoniou}, V.}, \bibinfo{author}{{Zezas}, A.},
  \bibinfo{author}{{Hatzidimitriou}, D.} \& \bibinfo{author}{{Kalogera}, V.}
\newblock \bibinfo{title}{{Star Formation History and X-ray Binary Populations:
  The Case of the Small Magellanic Cloud}}.
\newblock \emph{\bibinfo{journal}{\apjl}} \textbf{\bibinfo{volume}{716}},
  \bibinfo{pages}{L140--L145} (\bibinfo{year}{2010}).

\bibitem{2010MNRAS.401..252C}
\bibinfo{author}{{Coe}, M.~J.}, \bibinfo{author}{{McBride}, V.~A.} \&
  \bibinfo{author}{{Corbet}, R.~H.~D.}
\newblock \bibinfo{title}{{Exploring accretion theory with X-ray binaries in
  the Small Magellanic Cloud}}.
\newblock \emph{\bibinfo{journal}{\mnras}} \textbf{\bibinfo{volume}{401}},
  \bibinfo{pages}{252--256} (\bibinfo{year}{2010}).

\bibitem{2009ApJ...699.1573L}
\bibinfo{author}{{Linden}, T.}, \bibinfo{author}{{Sepinsky}, J.~F.},
  \bibinfo{author}{{Kalogera}, V.} \& \bibinfo{author}{{Belczynski}, K.}
\newblock \bibinfo{title}{{Probing Electron-Capture Supernovae: X-Ray Binaries
  in Starbursts}}.
\newblock \emph{\bibinfo{journal}{\apj}} \textbf{\bibinfo{volume}{699}},
  \bibinfo{pages}{1573--1577} (\bibinfo{year}{2009}).

\bibitem{1994AJ....108.2348A}
\bibinfo{author}{{Ashman}, K.~M.}, \bibinfo{author}{{Bird}, C.~M.} \&
  \bibinfo{author}{{Zepf}, S.~E.}
\newblock \bibinfo{title}{{Detecting bimodality in astronomical datasets}}.
\newblock \emph{\bibinfo{journal}{\aj}} \textbf{\bibinfo{volume}{108}},
  \bibinfo{pages}{2348--2361} (\bibinfo{year}{1994}).

\end{thebibliography}

\begin{thebibliography}{10}
\expandafter\ifx\csname url\endcsname\relax
  \def\url#1{\texttt{#1}}\fi
\expandafter\ifx\csname urlprefix\endcsname\relax\def\urlprefix{URL }\fi
\providecommand{\bibinfo}[2]{#2}
\providecommand{\eprint}[2][]{\url{#2}}

\bibitem[26]{1977Natur.266..123R}
\bibinfo{author}{{Rappaport}, S.} \& \bibinfo{author}{{Joss}, P.~C.}
\newblock \bibinfo{title}{{Binary X-ray pulsars}}.
\newblock \emph{\bibinfo{journal}{\nat}} \textbf{\bibinfo{volume}{266}},
  \bibinfo{pages}{123--125} (\bibinfo{year}{1977}).

\bibitem[27]{1982Ap&SS..81..379B}
\bibinfo{author}{{Bhatt}, H.~C.}
\newblock \bibinfo{title}{{The pulse-period distribution of binary X-ray
  pulsars}}.
\newblock \emph{\bibinfo{journal}{\apss}} \textbf{\bibinfo{volume}{81}},
  \bibinfo{pages}{379--385} (\bibinfo{year}{1982}).

%
\bibitem[28]{2010ApJ...718.1266M}
\bibinfo{author}{{Muratov}, A.~L.} \& \bibinfo{author}{{Gnedin}, O.~Y.}
\newblock \bibinfo{title}{{Modeling the Metallicity Distribution of Globular
  Clusters}}.
\newblock \emph{\bibinfo{journal}{\apj}} \textbf{\bibinfo{volume}{718}},
  \bibinfo{pages}{1266--1288} (\bibinfo{year}{2010}).
\newblock \eprint{1002.1325}.

\bibitem[29]{silverman}
\bibinfo{author}{Silverman, B.~W.}
\newblock \bibinfo{title}{{Using kernel density estimates to investigate
  multimodality}}.
\newblock \emph{\bibinfo{journal}{Journal of the Royal Statistical Society,
  Series B}} \textbf{\bibinfo{volume}{43}}, \bibinfo{pages}{97--99}
  (\bibinfo{year}{1981}).

\bibitem[30]{convex}
\bibinfo{author}{{Gonz\'{a}lez-Manteiga}, W.} \& \bibinfo{author}{{Cuevas}, A.}
\newblock \bibinfo{title}{{Data-driven smoothing based on convexity
  properties}}.
\newblock In \bibinfo{editor}{{G. Roussas}} (ed.)
  \emph{\bibinfo{booktitle}{Nonparametric functional estimation and related
  topics}}, vol. \bibinfo{volume}{335} of \emph{\bibinfo{series}{NATO ASI
  Series C}}, \bibinfo{pages}{225--240} (\bibinfo{year}{1991}).

\bibitem[31]{hall}
\bibinfo{author}{{Hall}, P.} \& \bibinfo{author}{{York}, M.}
\newblock \bibinfo{title}{{On the calibration of Silverman's test for
  multimodality}}.
\newblock \emph{\bibinfo{journal}{Statistica Sinica}}
  \textbf{\bibinfo{volume}{11}}, \bibinfo{pages}{515--536}
  (\bibinfo{year}{2001}).

\bibitem[32]{2008MNRAS.386.1426K}
\bibinfo{author}{{Knigge}, C.}, \bibinfo{author}{{Scaringi}, S.},
  \bibinfo{author}{{Goad}, M.~R.} \& \bibinfo{author}{{Cottis}, C.~E.}
\newblock \bibinfo{title}{{The intrinsic fraction of broad-absorption line
  quasars}}.
\newblock \emph{\bibinfo{journal}{\mnras}} \textbf{\bibinfo{volume}{386}},
  \bibinfo{pages}{1426--1435} (\bibinfo{year}{2008}).
\newblock \eprint{0802.3697}.

\bibitem[33]{wys}
\bibinfo{author}{Wyszomirski, T.}
\newblock \bibinfo{title}{{Detecting and displaying size bimodality: Kurtosis,
  skewness and bimodalizable distributions}}.
\newblock \emph{\bibinfo{journal}{Journal of Theoretical Biology}}
  \textbf{\bibinfo{volume}{158}}, \bibinfo{pages}{109--128}
  (\bibinfo{year}{1992}).

\bibitem[34]{2003A&A...405..677N}
\bibinfo{author}{{North}, P.} \& \bibinfo{author}{{Zahn}, J.}
\newblock \bibinfo{title}{{Circularization in B-type eclipsing binariesin the
  Magellanic Clouds}}.
\newblock \emph{\bibinfo{journal}{\aap}} \textbf{\bibinfo{volume}{405}},
  \bibinfo{pages}{677--684} (\bibinfo{year}{2003}).

\bibitem[35]{1995MNRAS.274..461B}
\bibinfo{author}{{Brandt}, N.} \& \bibinfo{author}{{Podsiadlowski}, P.}
\newblock \bibinfo{title}{{The effects of high-velocity supernova kicks on the
  orbital properties and sky distributions of neutron-star binaries}}.
\newblock \emph{\bibinfo{journal}{\mnras}} \textbf{\bibinfo{volume}{274}},
  \bibinfo{pages}{461--484} (\bibinfo{year}{1995}).

\bibitem[36]{2002ApJ...574..364P}
\bibinfo{author}{{Pfahl}, E.}, \bibinfo{author}{{Rappaport}, S.},
  \bibinfo{author}{{Podsiadlowski}, P.} \& \bibinfo{author}{{Spruit}, H.}
\newblock \bibinfo{title}{{A New Class of High-Mass X-Ray Binaries:
  Implications for Core Collapse and Neutron Star Recoil}}.
\newblock \emph{\bibinfo{journal}{\apj}} \textbf{\bibinfo{volume}{574}},
  \bibinfo{pages}{364--376} (\bibinfo{year}{2002}).

%
%
\bibitem[37]{2001ApJ...546..455D}
\bibinfo{author}{{Delgado-Mart{\'{\i}}}, H.}, \bibinfo{author}{{Levine},
  A.~M.}, \bibinfo{author}{{Pfahl}, E.} \& \bibinfo{author}{{Rappaport}, S.~A.}
\newblock \bibinfo{title}{{The Orbit of X Persei and Its Neutron Star
  Companion}}.
\newblock \emph{\bibinfo{journal}{\apj}} \textbf{\bibinfo{volume}{546}},
  \bibinfo{pages}{455--468} (\bibinfo{year}{2001}).
\newblock \eprint{arXiv:astro-ph/0004258}.

\bibitem[38]{2011A&A...527A...7S}
\bibinfo{author}{{Staubert}, R.} \emph{et~al.}
\newblock \bibinfo{title}{{Finding a 24-day orbital period for the X-ray binary
  1A 1118-616}}.
\newblock \emph{\bibinfo{journal}{\aap}} \textbf{\bibinfo{volume}{527}},
  \bibinfo{pages}{A7+} (\bibinfo{year}{2011}).
\newblock \eprint{1012.2459}.

\end{thebibliography}

\end{document}